\documentclass[pra,showpacs,uplatex]{revtex4}

\usepackage{amsmath}
\usepackage{bm}
\usepackage[dvipdfmx]{graphicx}
\usepackage[dvipdfmx]{color}
\usepackage[dvipdfmx]{graphicx}
\usepackage{float}
\usepackage{ulem}
\usepackage{color}

\begin{document}

\title{Dark matter signals on a laser interferometer}

\author{Satoshi Tsuchida}
\email[Satoshi Tsuchida: ]{tsuchida@gwv.hep.osaka-cu.ac.jp}
\address{Department of Physics, Osaka City University, Osaka, Osaka, 558-8585, Japan}

\author{Nobuyuki Kanda}
\address{Department of Physics, Osaka City University, Osaka, Osaka, 558-8585, Japan,
Nambu Yoichiro Institute of Theoretical and Experimental Physics (NITEP), Osaka City University, Osaka, Osaka, 558--8585, Japan}

\author{Yousuke Itoh}
\address{Department of Physics, Osaka City University, Osaka, Osaka, 558-8585, Japan,
Nambu Yoichiro Institute of Theoretical and Experimental Physics (NITEP), Osaka City University, Osaka, Osaka, 558--8585, Japan}

\author{Masaki Mori}
\affiliation{Department of Physical Sciences, Ritsumeikan University, Kusatsu, Shiga, 525-8577, Japan}

\date{\today}

\begin{flushright}
  NITEP 28
\end{flushright}

\begin{abstract}
WIMPs are promising dark matter candidates.
A WIMP occasionally collides with a mirror
equipped with interferometric gravitational wave detectors such as LIGO, Virgo, KAGRA and the Einstein Telescope (ET).
When WIMPs collide with a mirror of an interferometer, we expect that characteristic motions of
the pendulum and mirror are excited,
and those signals could be extracted by highly sophisticated sensors developed for
gravitational wave detection.
We analyze the motions of the pendulum and mirror, and estimate the detectability of these motions.
For the ``Thin-ET" detector,
the signal-to-noise ratio may be
$ 1.7 \left( \frac{ m_{\rm{DM}} }{ 100~\rm{GeV} } \right) $,
where $ m_{\rm{DM}} $ is the mass of a WIMP.
We may set a more strict upper limit on the cross section between a WIMP and a nucleon
than the limits obtained by other experiments so far
when $ m_{\rm{DM}} $ is approximately lower than 0.2~GeV.
We find an order-of-magnitude improvement in the upper limit around $ m_{\rm{DM}} = 0.2~{\rm{GeV}} $.
\end{abstract}



\maketitle

\section{Introduction}
\label{sec:intro}

The first direct detection of a gravitational wave (GW) event was achieved by LIGO
(Laser Interferometer Gravitational-Wave Observatory) in 2015~\cite{GW150914}.
To date, ten binary black hole mergers~\cite{GW150914, GW151226, GW170104, GW170814, GW170608, GWBBHs}
and one binary neutron star signal~\cite{GW170817}
were detected in the first and second LIGO/Virgo observing runs (O1, O2).
LIGO and Virgo started the third observing run (O3) in April 2019.
KAGRA, the first cryogenic underground GW observatory,
is now under construction in Japan~\cite{KAGRA1, KAGRA2, KAGRA3, KAGRA4}, and it is planned to join the O3 run.
In addition, the third-generation GW detectors such as
Einstein Telescope (ET)~\cite{ETcqrg} and Cosmic Explorer~\cite{CE} are being proposed.
As the sensitivities of the current-generation GW detectors are so high,
these detectors can be sensitive not only to GWs, but also to external agents.
Namely, GW detectors could extract signals caused by
dark matter particles colliding with a mirror equipped with interferometers.

Candidates for dark matter may be categorized into two types.
One is macroscopic matter, such as MACHOs (massive compact halo objects),
whereas the other is microscopic matter, such as WIMPs (weakly interacting massive particles).
WIMPs are believed to be good candidates for dark matter to explain the structure of the present Universe,
and have an extensive allowed mass range of about 0.1~GeV to 10~TeV.
Methods explored so far to hunt for WIMPs include
collider searches, indirect detections, and direct detections:
for details, see, e.g., Refs.~\cite{Gelmini,BiYui}.
To prove the existence of WIMPs, direct detections,
where one observes possible nuclear recoils after WIMP-nucleon elastic scattering,
would be the most suitable method.
The cross section between a WIMP and a nucleon is expected to be extremely small.
So far, a couple of research groups have reported positive signals~\cite{DAMA,CDMS,CoGeNT},
but the results are still controversial and
it seems still premature to claim the existence of a WIMP.

We propose a search method for WIMP signals
using laser interferometric gravitational wave detectors.
Possible dark matter signals on laser interferometers have been investigated
in several works~\cite{Kashlinsky2016, Aoki, Yamamoto, Guo}.
However, calculations of the signals caused by direct interaction between a WIMP and nucleons
in the mirror of interferometers have not been considered in the literature yet.

In this paper,
we solve equations of motion for the behavior of the pendulum and mirror
induced by a WIMP collision with the mirror,
and obtain the characteristic amplitude spectrum.
Then, we derive the signal-to-noise ratio by comparing the signals to the design sensitivity of each detector,
and set an upper limit on the cross section between a WIMP and a nucleon.

\section{Dark matter flux and event rate}
\label{sec:flux_er}

The dark matter flux, $ {\Phi}_{\rm{DM}} $, around the Earth is given as follows~\cite{Baudis2012}:
\begin{eqnarray}
  \label{eq:flux}
  {\Phi}_{\rm{DM}} &=& n_{\rm{DM}} \times \langle v \rangle
    = \frac{ {\rho}_{\rm{DM}} }{ m_{\rm{DM}} } \langle v \rangle  \nonumber \\
                   &\cong& 6.6 \times 10^{4}~{\rm{ cm^{-2}~s^{-1} }}
                       \left( \frac{ {\rho}_{\rm{DM}} }{ 0.3~{ \rm{ GeV / cm^{3} } } }  \right)
                       \left( \frac{ 100~{\rm{GeV}} }{ m_{\rm{DM}} } \right)
                       \left( \frac{ \langle v \rangle }{ 220~{ \rm{ km / s } } } \right),
\end{eqnarray}
where
$ n_{\rm{DM}} $ is the number density of WIMPs,
$ \langle v \rangle $ is the mean velocity of WIMPs,
$ {\rho}_{\rm{DM}} $ is the local dark matter density,
and $ m_{\rm{DM}} $ is the mass of WIMPs.
Using this flux,
we can estimate the event rate, $ R $, of WIMP collisions with nucleons near the Earth as follows:
\begin{eqnarray}
  \label{eq:everate}
  R &=& \frac{ N_{\rm{A}} }{ A } {\Phi}_{\rm{DM}} {\sigma}_{\rm{WN}} (A)  \nonumber \\
    &\cong& 0.13~\frac{ \rm{events} }{ \rm{kg \cdot year} }
        \left( \frac{ 100~{\rm{g/mol}} }{ A } \right)
        \left( \frac{ {\rho}_{\rm{DM}} }{ 0.3~{ \rm{ GeV / cm^{3} } } }  \right)
        \left( \frac{ 100~{\rm{GeV}} }{ m_{\rm{DM}} } \right)
        \left( \frac{ \langle v \rangle }{ 220~{ \rm{ km / s } } } \right)
        \left( \frac{ {\sigma}_{\rm{WN}} (A) }{ 10^{-38}~{\rm{cm^{2}}} } \right),
\end{eqnarray}
where
$ N_{\rm{A}} = 6.02 \times 10^{23}~{\rm{mol}}^{-1} $ is the Avogadro constant,
$ A $ is the molar mass of the target nucleus,
and
$ {\sigma}_{\rm{WN}} (A) $ is the cross section between a WIMP and a nucleon.
The value of the cross section may affect the lifetime of WIMPs;
thus, the evaluation of the cross section could play a important role to elucidate the nature of WIMPs.

\section{Expected Dark Matter Signals}

\begin{figure}[t]
  \begin{center}
    \includegraphics[width=120mm]{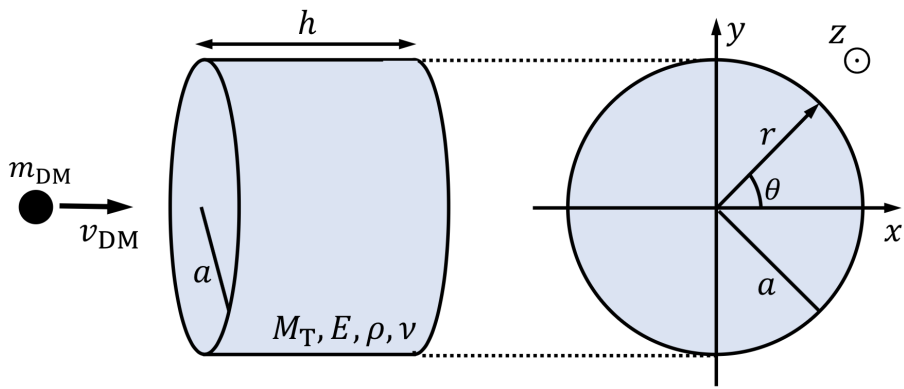}
    \caption{(Color Online).
    The schematic image for a collision of a WIMP with the mirror
    equipped with a laser interferometer.
 }
    \label{fig:situation}
  \end{center}
\end{figure}

The schematic image for a collision of a WIMP with the mirror
is shown in Fig.~\ref{fig:situation}.
The parameters $ M_{\rm{T}} $, $ E $, $ \rho $, $ \nu $, $ a $, and $ h $ in Fig.~\ref{fig:situation}
are the mass, Young's modulus, matter density, Poisson's ratio, radius,
and thickness of the mirror, respectively.
The values of these parameters for the detectors are given in Table~\ref{tab:mirror}.
When a WIMP collides with a nucleon in the mirror,
we expect that various characteristic motions of the pendulum and mirror occur.
In this paper, we consider the induced signals due to
(i) pendulum (translation) motion and
(ii) elastic oscillation of the mirror.
We do not consider other motions such as
the rotation of the mirror or the violin mode of the pendulum, and so on.
Here, we derive the expressions for signals due to (i) and (ii).

(i)
{\it{Pendulum (translation) motion}}:
First, we consider the translation of the mirror, namely the motion of the pendulum.
The equation of motion for this mode is given by
\begin{eqnarray}
  \label{eq:pen_eom}
  \frac{ d^{2} z_{\rm{Pend}} (t) }{ dt^{2} } + \frac{ 2 \pi f_{0} }{ Q_{\rm{P}} } \frac{ dz_{\rm{Pend}} (t) }{ dt } + \left( 2 \pi f_{0} \right)^{2} z_{\rm{Pend}} (t) = \frac{ F(t) }{ M_{\rm{T}} },
\end{eqnarray}
where $ Q_{\rm{P}} \sim 10^{7} $ is the quality factor,
$ f_{0} \simeq 1~{\rm{Hz}} $ is the resonance frequency of the pendulum,
and
$ F(t) $ is the external force given by a WIMP collision:
\begin{eqnarray}
  \label{eq:force_pen}
  F(t) = P_{\rm{DM}} \delta ( t ),
\end{eqnarray}
where $ P_{\rm{DM}} = m_{\rm{DM}} v_{\rm{DM}} $ is the momentum of a WIMP,
$ v_{\rm{DM}} = 220~{\rm{km/s}} $ is the typical velocity of WIMPs,
and
we assume the collision happens at $ t = 0 $.
Here, we assume the delta-functional force for $ F(t) $.
When a WIMP that has $ m_{\rm{DM}} = 100~{\rm{GeV}} $ collides with a nucleon in the mirror
and scatters elastically,
the nucleon will have a kinetic energy of about $ 30~{\rm{keV}} $.
This energy may be higher than the binding energy of intermolecular force in the mirror,
so the nucleon would give rise to a ``secondary" nucleon.
By using the SRIM (the Stopping and Range of Ions in Matter) calculation tool~\cite{SRIM},
we can show that the secondary nucleon may be stopped within about $ 10^{-12}~{\rm{s}} $,
and this timescale is much shorter than the sampling time of gravitational wave data acquisition systems.
Thus, we can ignore the effect of the secondary nucleon,
and we can approximately describe the collision
using a delta function as in Eq.~(\ref{eq:force_pen}).

The solution of the equation of motion~(\ref{eq:pen_eom}) is obtained as in a
damped sinusoidal waveform:
\begin{eqnarray}
  \label{eq:sol_pen}
  z_{\rm{Pend}} (t) = \frac{ P_{\rm{DM}} }{2 \pi M_{\rm{T}} f_{0} \sqrt{ 1 - \frac{ 1 }{ 4 Q_{\rm{P}}^{2} } } }
          \exp \left[ - \frac{ \pi f_{0} }{ Q_{\rm{P}} } t \right]
          \sin \left( 2 \pi f_{0} \sqrt{ 1 - \frac{ 1 }{ 4 Q_{\rm{P}}^{2} } } t \right).
\end{eqnarray}
Using Fourier transformation, defined as $ \tilde{z} (f) = \int_{ - \infty }^{ \infty } z(t) e^{ - 2 \pi ift } dt $,
this solution can be written in the frequency domain as follows:
\begin{eqnarray}
  \label{eq:spc_pen}
  \left| \tilde{z}_{\rm{Pend}} (f) \right|
    = \frac{ P_{\rm{DM}} }{ 4 {\pi}^{2} M_{\rm{T}} }
        \frac{ 1 }{ \sqrt{ \left( - f^{2} + f_{0}^{2} \right)^{2} + \left( \frac{ f f_{0} }{ Q_{\rm{P}} } \right)^{2} } }.
\end{eqnarray}
This expression shows that the signal caused by the motion of the pendulum has a sharp peak at the resonance frequency $ f = f_{0} $,
and the signal is proportional to $ f^{-2} $ at higher frequencies than $ f_{0} $.

\begin{table}[b]
  \caption{Characteristic quantities of the mirrors for the interferometers.}
  \label{tab:mirror}
    \begin{tabular}{c|ccc}
      \multicolumn{1}{c|}{ }               &   \multicolumn{3}{c}{Laser Interferometers} \\ \hline
                      { }                  &   KAGRA   &   LIGO, Virgo  & Einstein Telescope \\ \hline \hline
                    Material               &  Sapphire &  Fused Silica  &    Fused Silica    \\
         Molar mass, $ A~[\rm{g/mol}] $    &   101.96  &     60.08      &       60.08        \\
    Mirror Mass, $ M_{\rm{T}}~[\rm{kg}] $  &     23    &       40       &        200         \\
       Density, $ \rho~[\rm{g/cm^{3}}] $   &    4.00   &      2.20      &        2.20        \\
             Radius, $ a~[\rm{cm}] $       &     11    &      17.5      &        31          \\
           Thickness, $ h~[\rm{cm}] $      &     15    &       20       &        30          \\
       Young's modulus, $ E~[\rm{GPa}] $   &    400    &      72.6      &       72.6         \\
            Poisson's ratio, $ \nu $       &    0.3    &      0.16      &       0.16         \\ \hline \hline
    \end{tabular}
\end{table}

(ii)
{\it{Elastic oscillation of the mirror}}:
Second, we consider the elastic oscillation of the mirror induced by a WIMP collision with the mirror
that has a cylindrical shape.
The equation of motion is given by
\begin{eqnarray}
  \label{eq:eom_ela}
  \frac{ {\partial}^{2} z_{\rm{Elas}} (t, r, \theta ) }{ {\partial} t^{2} }
    + \frac{ 2 \pi f_{\rm{e}} }{ Q_{\rm{M}} } \frac{ {\partial} z_{\rm{Elas}} (t, r , \theta ) }{ {\partial} t }
    + {\mathcal{D}} \left( {\nabla}^{2} \right)^{2} z_{\rm{Elas}} (t, r , \theta ) = 0,
\end{eqnarray}
where
$ \mathcal{D} = \frac{ E h^{2} }{ 12 \rho ( 1 - {\nu}^{2} ) } $ is the flexural rigidity,
$ Q_{\rm{M}} \sim 10^{7} $ is the quality factor of the mirror,
$ f_{\rm{e}} $ is the eigenfrequency of the elastic oscillation,
and
$ {\nabla}^{2} $ is the two-dimensional Laplacian.
The solution of Eq.~(\ref{eq:eom_ela}) is given by
\begin{eqnarray}
  \label{eq:ela_solution}
  z_{\rm{Elas}} ( t, r, \theta )
  = \sum_{m=0}^{\infty} \cos ( m \theta ) \sum_{n=0}^{\infty}
    K_{mn} R_{mn} (r) \exp \left[ - \frac{ {\pi} f_{mn} }{ Q_{\rm{M}} } t \right]
    \sin \left( 2 {\pi} f_{mn} \sqrt{ 1 - \frac{ 1 }{ 4 Q_{\rm{M}}^{2} } } t \right),
\end{eqnarray}
where
$ m $ corresponds to the number of nodal diameters,
$ n $ is the number of nodal circles,
$ f_{mn} $ denotes the eigenfrequency for each mode,
$ K_{mn} $ is a numerical constant depending on initial conditions,
and
$ R_{mn} (r) $ is a function of $ r $, which will be given below.

As for the boundary condition,
we assume that the mirror is a completely free cylinder,
since the mirror is not clamped.
In this situation, at the edge of the circle of the cylinder, $ r = a $,
the bending moment $ M_{r} (r=a) $ and shearing force $ V_{r} (r=a) $ should be zero---that is,
they satisfy the following conditions:
\begin{eqnarray}
  \label{eq:ela_bound}
  \left. M_{r} (r) \right|_{r=a}
  &=&
  \left. \left[ \frac{ {\partial}^{2} z_{\rm{Elas}} }{ {\partial} r^{2} }
  + \nu \left( \frac{1}{r} \frac{ {\partial} z_{\rm{Elas}} }{ {\partial} r }
  + \frac{1}{ r^{2} } \frac{ {\partial}^{2} z_{\rm{Elas}} }{ {\partial} {\theta}^{2} } \right) \right] \right|_{r=a}
  = 0, \nonumber \\
  \left. V_{r} (r) \right|_{r=a}
  &=&
  \left. \left[ \frac{ {\partial} }{ {\partial} r } \left( {\nabla}^{2} z_{\rm{Elas}} \right)
  + \frac{ 1 - {\nu} }{ r } \frac{ {\partial} }{ {\partial} r } \left( \frac{1}{r}
  \frac{ {\partial} z_{\rm{Elas}} }{ {\partial} {\theta} } \right) \right] \right|_{r=a}
  = 0.
\end{eqnarray}
These boundary conditions lead to the eigenvalue equation
\begin{eqnarray}
  \label{eq:lambda_bound}
  \frac{ {\lambda}_{mn}^{2} J_{m} ({\lambda}_{mn}) + ( 1 - \nu ) \left[ {\lambda}_{mn} J_{m}' ({\lambda}_{mn}) - m^{2} J_{m} ({\lambda}_{mn}) \right] }
  { {\lambda}_{mn}^{2} I_{m} ({\lambda}_{mn}) - ( 1 - \nu ) \left[ {\lambda}_{mn} I_{m}' ({\lambda}_{mn}) - m^{2} I_{m} ({\lambda}_{mn}) \right] }
  =
  \frac{ {\lambda}_{mn}^{3} J_{m}' ({\lambda}_{mn}) + ( 1 - \nu ) m^{2} \left[ {\lambda}_{mn} J_{m}' ({\lambda}_{mn}) - J_{m} ({\lambda}_{mn}) \right] }
  { {\lambda}_{mn}^{3} I_{m}' ({\lambda}_{mn}) - ( 1 - \nu ) m^{2} \left[ {\lambda}_{mn} I_{m}' ({\lambda}_{mn}) - I_{m} ({\lambda}_{mn}) \right] },
\end{eqnarray}
where
$ {\lambda}_{mn} = {\Omega}_{mn} a $, $ {\Omega}_{mn}^{4} = \frac{ ( 2 \pi f_{mn} )^{2} }{ {\mathcal{D}} } $,
$ J_{m} ({\lambda}) $ is a Bessel function,
$ I_{m} ({\lambda}) $ is a modified Bessel function,
$ J_{m}' ({\lambda}) = \frac{ {\partial} J_{m} ({\lambda}) }{ {\partial} {\lambda} } $,
and
$ I_{m}' ({\lambda}) = \frac{ {\partial} I_{m} ({\lambda}) }{ {\partial} {\lambda} } $.
From these relations, we obtain the eigenfrequency for each mode,
and these frequencies are listed in Table~\ref{tab:lambda}.
As can be expected, the eigenfrequency of each mode is smaller for a softer and thinner mirror.
The material of the mirrors equipped with the KAGRA is sapphire,
which is harder than the fused silica that constitutes the mirrors of LIGO and Virgo,
so the eigenfrequency of each mode for KAGRA is higher than that for the other mirrors.
On the other hand, the mirrors for ET are relatively thinner than the mirrors for
KAGRA, LIGO and Virgo;
thus, the mirrors for ET have a lower eigenfrequency for each mode.
\begin{table}[b]
  \caption{The value of eigenfrequency in units of $ \times 10^{4} $ [Hz] for each $ m $ and $ n $ for KAGRA (LIGO, Virgo) [ET].}
  \label{tab:lambda}
    \begin{tabular}{c|r|r|r|r|r|r}
    \multicolumn{1}{c|}{ } & \multicolumn{1}{c|}{$ m = 0 $} & \multicolumn{1}{c|}{$ m = 1 $} & \multicolumn{1}{c|}{$ m = 2 $} & \multicolumn{1}{c|}{$ m = 3 $} & \multicolumn{1}{c|}{$ m = 4 $} & \multicolumn{1}{c}{$ m = 5 $} \\ \hline \hline
           $ n = 0 $       &   \multicolumn{1}{c|}{ --- }   &   \multicolumn{1}{c|}{ --- }   &       3.20 (1.01) [0.481]      &       7.43 (2.31) [1.10]       &       13.0 (4.03) [1.93]       &      20.0 (6.15) [2.94]       \\
           $ n = 1 $       &       5.38 (1.51) [0.724]      &       12.2 (3.54) [1.69]       &       21.1 (6.17) [2.95]       &       31.6 (9.33) [4.46]       &       43.9 (13.0) [6.20]       &      57.8 (17.1) [8.17]       \\
           $ n = 2 $       &       23.0 (6.66) [3.18]       &       35.7 (10.4) [4.97]       &       50.4 (14.7) [7.04]       &       66.8 (19.6) [9.36]       &       85.0 (25.0) [11.9]       &      105  (30.8) [14.7]       \\
           $ n = 3 $       &       52.4 (15.3) [7.30]       &       71.0 (20.7) [9.91]       &       91.5 (26.8) [12.8]       &       114  (33.3) [15.9]       &       138  (40.4) [19.3]       &      164  (48.0) [22.9]       \\
           $ n = 4 $       &       93.6 (27.3) [13.1]       &       118  (34.5) [16.5]       &       145  (42.2) [20.2]       &       173  (50.5) [24.1]       &       203  (59.3) [28.3]       &      234  (68.6) [32.8]       \\
           $ n = 5 $       &       147  (42.8) [20.5]       &       177  (51.7) [24.7]       &       209  (61.2) [29.2]       &       243  (71.2) [34.0]       &       279  (81.7) [39.0]       &      317  (92.7) [44.3]       \\ \hline
    \end{tabular}
\end{table}

Then, we derive the displacement of the mirror and function $ R_{mn} (r) $
by using Eqs.~(\ref{eq:ela_bound}) and (\ref{eq:lambda_bound}),
so the solution of Eq.~(\ref{eq:eom_ela}) is written as
\begin{eqnarray}
  \label{eq:solution}
  z_{\rm{Elas}} ( t , r , \theta ) &=&
  \sum_{m=0}^{\infty} \cos ( m \theta ) \sum_{n=0}^{\infty}
  K_{mn} R_{mn} (r) \exp \left[ - \frac{ {\pi} f_{mn} }{ Q_{\rm{M}} } t \right]
  \sin \left( 2 {\pi} f_{mn} \sqrt{ 1 - \frac{ 1 }{ 4 Q_{\rm{M}}^{2} } } t \right), \nonumber \\
  {\rm{with}} \ \
  R_{mn} (r) &=& \left[ J_{m} \left( {\Omega}_{mn} r \right) +
  \frac{ {\lambda}_{mn}^{3} J_{m}' ({\lambda}_{mn}) + ( 1 - \nu ) m^{2} \left[ {\lambda}_{mn} J_{m}' ({\lambda}_{mn}) - J_{m} ({\lambda}_{mn}) \right] }{ {\lambda}_{mn}^{3} I_{m}' ({\lambda}_{mn}) - ( 1 - \nu ) m^{2} \left[ {\lambda}_{mn} I_{m}' ({\lambda}_{mn}) - I_{m} ({\lambda}_{mn}) \right] }
  I_{m} \left( {\Omega}_{mn} r \right) \right].
\end{eqnarray}
Using Fourier transformation, we obtain the displacement in the frequency domain as
\begin{eqnarray}
  \label{eq:pow_spe_ela}
  \left| \tilde{z}_{\rm{Elas}} ( f , r , \theta ) \right|
  &\simeq&
  \frac{1}{2 \pi} \sqrt{ 1 - \frac{ 1 }{ 4 Q_{\rm{M}}^{2} } }
  \sum_{m=0}^{\infty} \cos ( m \theta )
  \sum_{n=0}^{\infty} K_{mn} f_{mn} R_{mn} (r)
  \frac{ 1 }{ \sqrt{ \left( - f^{2} + f_{mn}^{2} \right)^{2} + \left( \frac{ f f_{mn} }{ Q_{\rm{M}} } \right)^{2} } }.
\end{eqnarray}
Thus, the signal caused by elastic oscillation also has sharp peaks at the resonance frequencies $ f = f_{mn} $.
To calculate $ K_{mn} $, we consider the momentum conservation law that is given by
\begin{eqnarray}
  \label{eq:mom_con}
  P_{\rm{DM}} \delta \left( \boldsymbol{r} - \boldsymbol{r}_{0} \right)
    = 2 \pi \rho h \sum_{m=0}^{\infty} \cos ( m \theta ) \sum_{n=0}^{\infty} K_{mn} f_{mn} R_{mn} (r),
\end{eqnarray}
where
$ {\boldsymbol{r}}_{0} = ( r_{0}, {\theta}_{0} ) $ means the collision point of the WIMP on the mirror.
We multiply $ R_{pq} (r) \cos (p \theta) $ for both sides,
and integrate over the entire region of the mirror surface,
obtaining
\begin{eqnarray}
  \label{eq:mom_eq}
  P_{\rm{DM}} R_{mn} (r_{0}) \cos ( m \theta_{0} )
  =
  2 \pi \rho h K_{mn} f_{mn} \int_{0}^{a} R_{mn}^{2} (r) r dr \int_{0}^{2 \pi} \cos^{2} (m \theta ) d {\theta}.
\end{eqnarray}

We note that
the modes that contribute to the displacement at the center of the circle of the mirror
should play a key role in evaluating the effects of the signals caused by a WIMP collision,
since laser beams used for measuring the differential displacement of the arm length
irradiate the center of the circle of the mirror.
Thus, hereafter, we only consider the elastic oscillations at the center of the circle
that correspond to $ m = 0 $ modes.

We derive the numerical factor $ K_{0n} $ for each $ n $ mode as follows:
\begin{eqnarray}
  \label{eq:k_0n}
  K_{0n} ( r_{0} )
  =
  \frac{ P_{\rm{DM}} R_{0n} (r_{0}) }{ 4 {\pi}^{2} \rho h f_{0n} \int_{0}^{a} R_{0n}^{2} (r) r dr }.
\end{eqnarray}
Using $ K_{0n} $,
we obtain the magnitude of the displacement
at $ f = f_{0n} $ and $ r = 0 $
for each $ n $ mode and $ r_{0} $ as follows:
\begin{eqnarray}
  \label{eq:amp_elas_0n}
  \vert \tilde{z}_{\rm{Elas}} (f=f_{0n},r=0) \vert = \left| \frac{ 1 }{ 2 \pi } \sqrt{ 1 - \frac{1}{ 4 Q_{\rm{M}}^{2} } }
  K_{0n} (r_{0}) R_{0n} (r=0) Q_{\rm{M}} \frac{1}{ f_{0n} } \right|.
\end{eqnarray}
The values of them
are summarized in Table~\ref{tab:amp} with $ m_{\rm{DM}} = 100~{\rm{GeV}} $.
When a WIMP collides with the mirror at the center of the circle ($ r_{0} = 0 $),
the displacement $ \vert \tilde{z}_{\rm{Elas}} (f=f_{0n},r=0) \vert $ attains the maximum for each $ n $ mode.

\begin{table}[b]
  \caption{The magnitude of the displacement $ \vert \tilde{z}_{\rm{Elas}} (f=f_{0n},r=0) \vert $ ($ \times 10^{-26} $) for KAGRA (LIGO, Virgo) [ET] with $ m_{\rm{DM}} = 100~{\rm{GeV}} $.}
  \label{tab:amp}
  \begin{tabular}{c|c|c|c|c|c|c} \hline
    \multicolumn{1}{c|}{} & \multicolumn{6}{c}{The collision point of the WIMP on the mirror, $ r_{0} $} \\ \hline \hline
        { }   &      $ 0.0 a $     &     $ 0.1 a $      &     $ 0.2 a $      &     $ 0.3 a $      &     $ 0.4 a $      &     $ 0.5 a $      \\ \hline
    $ n = 1 $ & 60.8 (404)  [376]  & 59.2 (394)  [366]  & 54.4 (363)  [337]  & 46.8 (312)  [290]  & 36.7 (246)  [229]  & 24.6 (166)  [155]  \\
    $ n = 2 $ & 8.19 (52.4) [48.8] & 7.42 (47.6) [44.2] & 5.35 (34.4) [32.0] & 2.54 (16.5) [15.4] & 0.23 (1.19) [1.11] & 2.25 (14.2) [13.2] \\
    $ n = 3 $ & 2.34 (14.9) [13.8] & 1.86 (11.8) [11.0] & 0.70 (4.44) [4.13] & 0.44 (2.79) [2.60] & 0.94 (5.98) [5.56] & 0.65 (4.15) [3.85] \\
    $ n = 4 $ & 0.98 (6.19) [5.76] & 0.63 (4.00) [3.72] & 0.05 (0.31) [0.28] & 0.39 (2.49) [2.31] & 0.17 (1.09) [1.01] & 0.21 (1.33) [1.24] \\
    $ n = 5 $ & 0.50 (3.15) [2.93] & 0.24 (1.50) [1.39] & 0.15 (0.95) [0.88] & 0.13 (0.85) [0.79] & 0.11 (0.68) [0.63] & 0.10 (0.66) [0.61] \\ \hline \hline
        { }   &     $ 0.6 a $      &     $ 0.7 a $      &     $ 0.8 a $      &     $ 0.9 a $      &     $ 1.0 a $      &        ---         \\ \hline
    $ n = 1 $ & 11.3 (77.4) [72.0] & 2.87 (17.3) [16.1] & 17.2 (114)  [106]  & 31.4 (211)  [197]  & 45.1 (307)  [285]  &                    \\
    $ n = 2 $ & 3.04 (19.4) [18.0] & 2.50 (16.1) [14.9] & 0.88 (5.83) [5.42] & 1.35 (8.44) [7.85] & 3.72 (23.9) [22.2] &                    \\
    $ n = 3 $ & 0.06 (0.36) [0.34] & 0.60 (3.75) [3.49] & 0.55 (3.48) [3.23] & 0.05 (0.28) [0.26] & 0.87 (5.50) [5.11] &                    \\
    $ n = 4 $ & 0.26 (1.65) [1.54] & 0.02 (0.14) [0.13] & 0.23 (1.43) [1.33] & 0.07 (0.43) [0.40] & 0.31 (1.98) [1.84] &                    \\
    $ n = 5 $ & 0.09 (0.56) [0.52] & 0.09 (0.56) [0.52] & 0.07 (0.46) [0.42] & 0.06 (0.39) [0.36] & 0.14 (0.90) [0.84] &                    \\ \hline
  \end{tabular}
\end{table}

\section{Limit on the Cross Section between a WIMP and a Nucleon}

Here, we calculate the signal-to-noise ratio (SNR) $ \varrho $
and estimate the upper limit on the cross section between a WIMP and a nucleon $ {\sigma}_{\rm{WN}} $.
To calculate the SNR, we introduce the characteristic amplitude spectrum $ \sqrt{ S_{a} (f) } $
that is defined by
\begin{eqnarray}
  \label{eq:chara_spec}
  \sqrt{ S_{a} (f) } = \sqrt{ 4 f \frac{ \vert \tilde{z} (f) \vert^{2} }{ L^{2} } },
\end{eqnarray}
where the square modulus of the amplitude $ \vert \tilde{z} (f) \vert^{2} $ is given by
$ \vert \tilde{z} (f) \vert^{2} = \vert \tilde{z}_{\rm{Pend}} (f) \vert^{2} + \vert \tilde{z}_{\rm{Elas}} (f) \vert^{2} $,
and $ L $ is the arm length of an interferometer.
Using the spectrum $ \sqrt{ S_{a} (f) } $, the SNR is given by
\begin{eqnarray}
  \label{eq:snr_def}
  {\varrho}^{2}
  &=&
  \int_{ f_{\rm{min}} }^{ f_{\rm{max}} } \frac{ S_{a} (f) }{ S_{n} (f) } \frac{df}{f},
\end{eqnarray}
where $ S_{n} (f) $ is the one-sided power spectral density of the detector in consideration
that includes resident and possible backgrounds such as a seismic noise, radiation pressure noise,
shot noise, thermal noise, etc.
Here $ f_{\rm{min}} $ and $ f_{\rm{max}} $ are the minimum and the maximum frequencies
of the design sensitivity curves for the detectors given in Refs.~\cite{LIGOsens,KAGRAsens}.
As mentioned above, the signal spectrum $ S_{a} (f) $ has sharp peaks at
the eigenfrequencies and small values for other frequency regions,
so the contributions of the peaks predominantly increase the SNR.
However, most of the eigenfrequencies for KAGRA, LIGO, Virgo, and ET are
outside of the sensitivity curves for the detectors;
thus the SNR cannot attain enough values to detect these signals.

\begin{figure}[t]
  \begin{center}
    \vspace{20mm}
    \hspace{-45mm}
    \includegraphics[width=140mm]{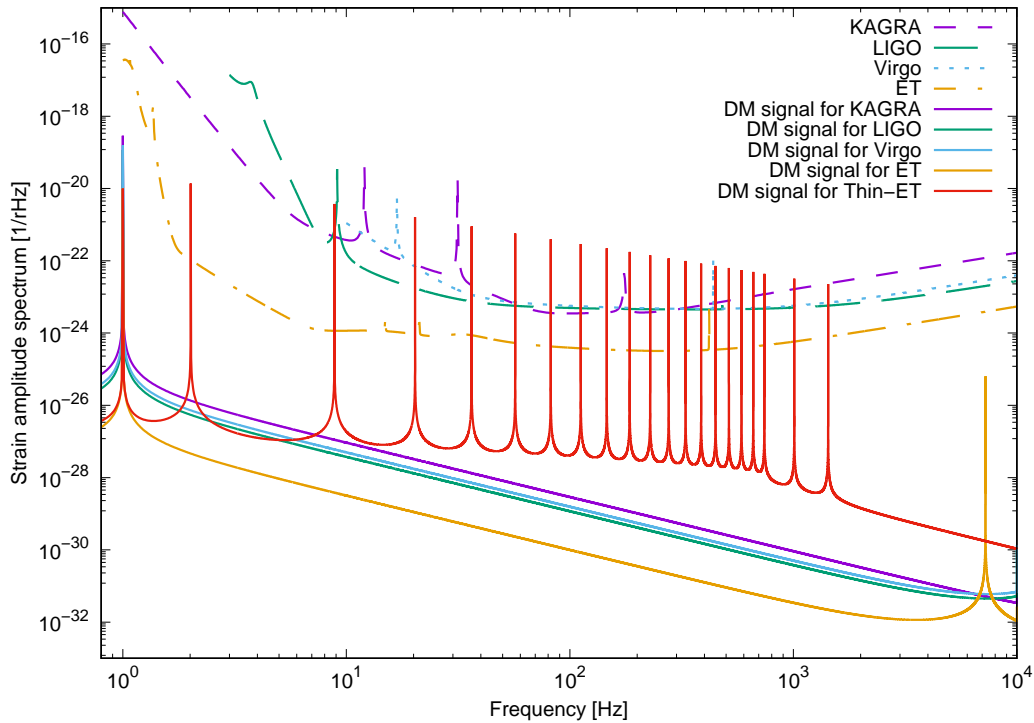}
    \vspace{-20mm}
    \caption{(Color Online).
    The characteristic amplitude spectra $ \sqrt{ S_{a} (f) } $ and design sensitivities for
    the existing or planned interferometers dedicated for gravitational wave observations
    \cite{LIGOsens,KAGRAsens}.
 }
    \label{fig:chara_spec}
  \end{center}
\end{figure}

Alternatively, we can propose a ``Thin-ET" detector to extract the signal caused by a WIMP collision.
Mirrors of the Thin-ET detector would have thinner thickness ($ h = 0.5~{\rm{cm}} $)
and larger radius ($ a = 240~{\rm{cm}}$),
and the other parameters of the mirrors and the arm length
are the same as those of the ET detector.
Thus, the sensitivity curve of the Thin-ET detector would be the same as
that of ET by using the calculation in Ref.~\cite{CA}.
Since the thin-thickness and large-radius cylinder has low eigenfrequencies,
the many sharp peaks can be in the observation frequency band.
Thus, we expect that we can obtain a larger SNR for the Thin-ET detector than the SNRs for other interferometers.
The characteristic amplitude spectra $ \sqrt{ S_{a} (f) } $ and design sensitivities for
the existing or planned interferometers dedicated for gravitational wave observations
are shown in Fig.~\ref{fig:chara_spec}.
This figure indicates that most of the peak magnitudes at the eigenfrequencies for the Thin-ET detector
may be higher than the given sensitivity curve,
so we expect that the Thin-ET detector has a moderate SNR value.
From the above calculation, the SNR is proportional to mass of a WIMP,
so we can write the SNR as $ {\varrho} = {\varrho}_{\rm{fact}} \left( \frac{ m_{\rm{DM}} }{ 100~{\rm{GeV}} } \right) $,
where $ {\varrho}_{\rm{fact}} \simeq 1.7 $ for the Thin-ET detector.

Since we know the expected waveform of the dark matter signal considered in this paper,
it is most optimal to detect the signal using a detection statistic based on the matched filtering technique,
which is widely used in the gravitational wave data analysis community.
We declare signal detection if our detection statistic exceeds, say, $ 5 \sigma $.
If not, we conclude no detection and
proceed to set an upper limit on the cross section between a WIMP and a nucleon.

\begin{figure}[t]
  \begin{center}
    \includegraphics[width=140mm]{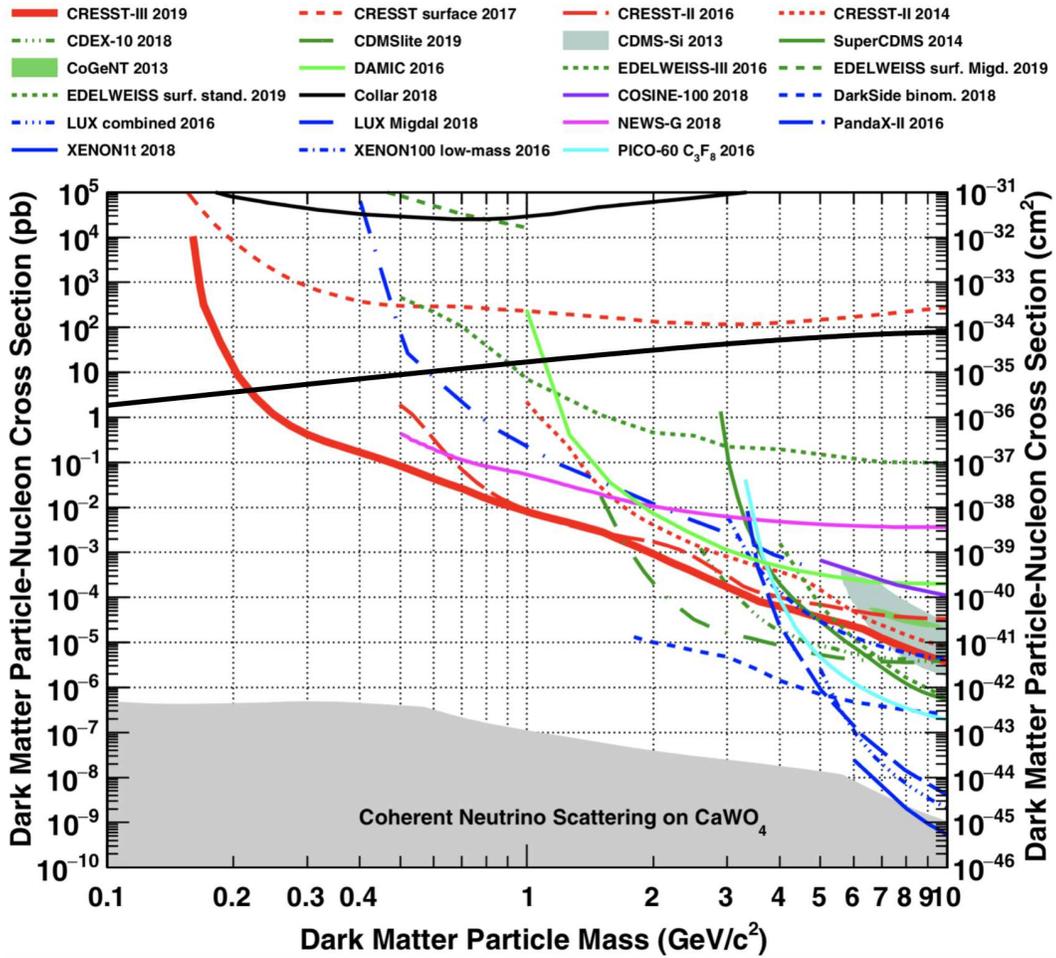}
    \caption{(Color Online).
    Upper limits on the cross section $ {\sigma}_{\rm{WN}} $ obtained by
    our calculation for the Thin-ET detector (thick solid black line)
    superimposed on Fig. 7 in Ref.~\cite{CRESST}.
    Detailed discussions for other experimental results obtained so far are given in Refs.~\cite{CSfig1,CSfig2,CSfig3,CSfig4,CSfig5,CSfig6,CSfig7,CSfig8,CSfig9,CSfig10,
    CSfig11,CSfig12,CSfig13,CSfig14,CSfig15,CSfig16,CSfig17,CSfig18,CSfig19,CSfig20}.
 }
    \label{fig:cross_section}
  \end{center}
\end{figure}

The number of collision events follows a Poisson distribution with the expected number of events $ \lambda $
given by $ \lambda \equiv \epsilon M_{\rm{T}} R T_{\rm{obs}} $,
where $ \epsilon $ is the detection efficiency and $ T_{\rm{obs}} $ is the observation time.
The detection efficiency may be calculated based on the detection threshold on our detection statistic ($ 5 \sigma $),
the expected signal-to-noise ratio given by Eq.~(\ref{eq:snr_def}), and the statistical property of detector noise.
We assume that the noise of a laser interferometric gravitational wave detector
follows a stationary Gaussian distribution, which is a good approximation to the first order.
The upper limit on the event rate at a 90{\%} confidence level, $ R_{90} $, may then be calculated using
\begin{eqnarray}
  \label{eq:er90}
  R_{90} = \frac{ 2.303 }{ {\epsilon} M_{\rm{T}} T_{\rm{obs}} }.
\end{eqnarray}
Using Eq.~(\ref{eq:er90}), we obtain the upper limit on the cross section $ {\sigma}_{\rm{WN}} $ as follows:
\begin{eqnarray}
  \label{eq:limit_cs}
  {\sigma}_{\rm{WN}}
  \simeq
  \frac{ 8.9 }{ \epsilon } \times 10^{-40}~{\rm{cm^{2}}}
  \left( \frac{ 200~\rm{kg} }{ M_{\rm{T}} } \right)
  \left( \frac{ 1~\rm{year} }{ T_{\rm{obs}} } \right)
  \left( \frac{ A }{100~{\rm{g/mol}}} \right)
  \left( \frac{ m_{\rm{DM}}}{ 100~{\rm{GeV}} } \right)
  \left( \frac{ 240~{\rm{cm}} }{ a } \right)
  \left( \frac{ a + h }{ 240.5~{\rm{cm}} } \right),
\end{eqnarray}
where the local dark matter density and the mean velocity of WIMPs
are fixed at $ {\rho}_{\rm{DM}} = 0.3~{\rm{GeV/cm^{3}}} $ and
$ \langle v \rangle = 220~{\rm{km/s}} $, respectively.
The last two factors in Eq.~(\ref{eq:limit_cs}) mean
the ratio between the surface area of two bottom faces and the total surface area of the mirror.
Our possible upper limit on the cross section as a function of the WIMP mass,
along with those by other experiments, is shown in Fig.~\ref{fig:cross_section}.
This figure implies that, in the low-WIMP-mass region ($ \lesssim 0.2~{\rm{GeV}} $),
we could set more strict upper limits on the cross section than the limits obtained so far.
When the mass of the WIMP is just a little smaller than 0.2~GeV,
the upper limit would be improved by roughly an order of magnitude.

We note that
we should consider the effects of instrumental noises and have to distinguish target signals from these noises
when we analyze real data obtained by interferometric gravitational wave detectors.
Possible sources of such noises include
the collisions of ambient particles with a mirror and thermal fluctuation of the mirror.
The process to estimate the effect of the former noise will be given in Appendix~\ref{sec:app_rate}.
As a result, the magnitude of the strain equivalent noise amplitude may be
$ \sim 10^{-22} \times f^{-2} ~{\rm{ / \sqrt{Hz} }} $.
This scale is smaller than other noises, which are already included in $ S_{n} (f) $ of Eq.~(\ref{eq:snr_def}),
as can be seen from Fig.~{\ref{fig:chara_spec}};
thus, the noise induced by the collisions of ambient particles with the mirror
does not have a critical influence on our analysis.
On the other hand, the latter noise would excite eigenmodes with eigenfrequencies
that are the same as the peak frequencies induced by collisions of the dark matter.
However, the ``effective temperature", which characterizes the effect of thermal noise,
can be lowered to
$ \sim Q^{-1} $ by applying a filter that has an optimal time length.
Thus, by using the filter,
we may be able to distinguish the dark matter signals from the thermal noise,
and we can extract the signals efficiently.
Cosmic rays give rise to signals similar to what we consider in this paper,
but the energy scale is sufficiently smaller than the sensitivity of the detector,
as is discussed in Ref.~\cite{Yamamoto}.
Thus, it may not affect our analysis.
Detailed discussions and estimations including such instrumental noises
and other possible motions of the pendulum and mirror
would be considered in future works.

\section{Conclusion}
\label{sec:conclusion}

When dark matter particles, such as WIMPs, collide with a mirror equipped with interferometers,
the motion of a pendulum and the elastic oscillation of the mirror are excited.
We performed a mode analysis of possible signals caused by a WIMP collision with the mirror
and calculated the signal-to-noise ratio, considering the design sensitivities of
the existing or planned detectors and the Thin-ET detector.
We derived that the signal-to-noise ratio may be
$ 1.7 \left( \frac{ m_{\rm{DM}} }{ 100~\rm{GeV} } \right) $
for the Thin-ET detector, and
we then estimated the upper limit on the cross section between a WIMP and a nucleon.
Such a Thin-ET detector enables us to set more strict upper limits on the cross section
in the low-WIMP-mass region ($ \lesssim 0.2~{\rm{GeV}} $)
that has never been explored before.
The limit would be improved by an order of magnitude around $ m_{\rm{DM}} = 0.2~{\rm{GeV}} $.

\appendix*

\section{Estimation for an effect of ambient particles}
\label{sec:app_rate}

Optical components of a laser interferometric gravitational wave detector
are in an ultrahigh vacuum of $ 10^{-9} $ torr.
The collisions of ambient particles with a mirror may frequently occur and become
a noise for the target signals.
Here, we estimate the effect of this noise as follows.

The collision rate of the ambient particles, $ R $, is given by
\begin{eqnarray}
  \label{eq:rate}
  R = n \cdot \bar{v} \cdot S,
\end{eqnarray}
where $ n = \frac{ P }{ k_{B} T } $ is the number density,
$ \bar{v} = \sqrt{ \frac{ 3 k_{B} T }{ m } } $ denotes the mean velocity,
$ m $ is the mass of the ambient particles,
$ P $ is the air pressure, $ T $ is the temperature in the vacuum chamber,
$ k_{B} $ is the Boltzmann constant,
and $ S = 2 \pi a^{2} $ corresponds to the surface area of two bottom faces of a mirror.
By substituting characteristic values for these parameters, we can estimate the collision rate as
\begin{eqnarray}
  \label{eq:rate_sec}
  R
  =
  P S \sqrt{ \frac{ 3 }{ m k_{B} T } }
  \cong
  3.0 \times 10^{15} ~{\rm{s^{-1}}}~
  \left( \frac{ P }{ 10^{-9}~{ \rm{Torr}} } \right) \left( \frac{ a }{ 0.175~{\rm{m}} } \right)^{2}
  \left( \frac{ 28 ~{\rm{GeV}} }{ m } \right)^{1/2} \left( \frac{ 300 ~{\rm{K}} }{ T } \right)^{1/2}.
\end{eqnarray}
An impulse received by each collision of a particle, $ I $, is given by
\begin{eqnarray}
  \label{eq:impulse}
  I = 2 m \bar{v}
    = 2 \sqrt{ 3 m k_{B} T }
    \cong 5.0 \times 10^{-23} ~{\rm{kg \cdot m \cdot s^{-1}}}
    \left( \frac{ m }{ 28 ~{\rm{GeV}} } \right)^{1/2} \left( \frac{ T }{ 300 ~{\rm{K}} } \right)^{1/2}.
\end{eqnarray}

We assume that the number of collisions of the ambient particles with the mirror follows the Poisson distribution,
and then the fluctuation of the rate, $ \delta R $, can be expressed as $ \delta R = \sqrt{R} $.
Thus, the strain equivalent noise amplitude induced by the collisions of ambient particles,
$ S_{\rm{amb}} (f) $, is given by
\begin{eqnarray}
  \! \! \! \!
  S_{\rm{amb}} (f)
  &=&
  \frac{ I \sqrt{R} }{ M _{\rm{T}} ( 2 \pi f )^{2} L }  \nonumber \\
  &\simeq&
  \frac{ 5.4 \times 10^{-22} }{ f^{2} } ~{\rm{ Hz^{-1/2} }}~
  \left( \frac{ P }{ 10^{-9}~{ \rm{Torr}} } \right)^{1/2} \left( \frac{ a }{ 0.175~{\rm{m}} } \right)
  \left( \frac{ m }{ 28 ~{\rm{GeV}} } \right)^{1/4} \left( \frac{ T }{ 300 ~{\rm{K}} } \right)^{1/4}
  \left( \frac{ 40~{\rm{kg}} }{ M_{\rm{T}} } \right) \left( \frac{ 3 ~{\rm{km}} }{ L } \right) \! . \ \ \ \ \
\end{eqnarray}
For the Thin-ET detector, $ S_{\rm{amb}} (f) $ becomes $ \sim 4 \times 10^{-22} f^{-2} ~{\rm{ Hz^{-1/2} }} $,
and this scale is smaller than other noises, as can be seen from Fig.~{\ref{fig:chara_spec}}.

\end{document}